\documentclass[aps,pra,onecolumn,groupedaddress,showpacs,showkeys,nofootonbib,preprint]{revtex4}

\usepackage{amsfonts}
\usepackage{amsmath}
\usepackage{amssymb}
\usepackage{graphicx}
\usepackage{epsfig}
\usepackage{amscd}
\usepackage{amsthm}
\usepackage{color}

\newtheorem{theor}{Theorem}[section]
\newtheorem{defin}{Definition}[section]
\newtheorem{lemma}{Lemma}[section]
\newtheorem{corol}{Corollary}[section]

\newcommand{\ket}[1]{|#1\rangle} 
\newcommand{\bra}[1]{\langle#1|}

\allowdisplaybreaks

\begin{document}

\title{A simple proof of the Jamio\l kowski criterion for complete positivity of linear maps}

\author{D. Salgado}
\email[]{david.salgado@uam.es}
\affiliation{Dpto. F\'{\i}sica Te\'{o}rica, Universidad Aut\'{o}noma de Madrid\\
28049 Cantoblanco, Madrid (Spain)}

\author{J.L. S\'{a}nchez-G\'{o}mez}
\email[]{jl.sanchezgomez@uam.es}
\altaffiliation{Permanent Address}
\affiliation{Dpto. F\'{\i}sica Te\'{o}rica, Universidad Aut\'{o}noma de Madrid\\
28049 Cantoblanco, Madrid (Spain)}

\author{M. Ferrero}
\email[]{ferrero@pinon.ccu.uniovi.es}
\altaffiliation{Permanent Address}
\affiliation{Dpto. F\'{\i}sica, Universidad de Oviedo\\
33007 Oviedo (Spain)}


\date{\today}

\begin{abstract}
We give a simple direct proof of the Jamio\l kowski criterion to check whether a linear map between matrix algebras is completely positive or not. This proof is more accesible for physicists than others found in the literature and provides a systematic method to give \emph{any} set of Kraus matrices of its Kraus decomposition.
\end{abstract}


\maketitle

\section{Introduction}

Complete positivity for linear maps defined upon operator algebras is a mathematical property which arises as a natural generalization of positivity of linear functionals on vector spaces \cite{Sti55a,Sto63a,Arv69a,Cho82a}. In Quantum Physics, it has been playing a remarkable role in several disciplines, especially in the analysis of entanglement and separability of composite systems (cf.\ \cite{Per96a,HorHorHor96a,LewBruCirKraKusSamSanTar00a,BruCirHorHulKraLewSan01a} and references therein), in the dynamics of open systems \cite{Dav76a,Lin76a,GorKosSud76a,AliLen87a,BrePet02a,BucHor02a} and in the Foundations of Quantum Mechanics \cite{Kra71a,Kra83a}. This property is used in different ways in these contexts, but in any case it is vital to have criteria to check whether a given linear map between operator algebras is completely positive (CP hereafter) or not.\\ 
Here we focus on a concrete one based on the Jamio\l kowski isomorphism \cite{Pil67a,Jam72a} and in particular we give a very accesible proof of this criterion for matrix algebras (i.e.\ for finite quantum systems) which, in the positive case, allows us to find \emph{any} set of Kraus operators of its Kraus decomposition \cite{Kra71a,Cho75a}. A barely accesible for physicists and more involved proof of this criterion can be found in \cite{Cho72a}.\\
The paper is divided as follows. In section \ref{Remin} we give a brief reminder of complete positivity (definition and main properties); in section \ref{Matrices} we provide the above mentioned proof of the Jamio\l kowski criterion, which is illustrated with useful and well-known examples in section \ref{Exa}. We end the paper in section \ref{Conclu} with a few conclusions.

\section{Complete Positivity: a brief reminder}
\label{Remin}

The rigorous and abstract definition of complete positivity is the following.

\begin{defin}
Let $\mathcal{A}$ be a $C^{*}$ algebra and $\alpha:\mathcal{A}\to\mathcal{A}$ an endomorphism. $\alpha$ is CP if the endomorphism $\alpha_{N}\equiv\alpha\otimes I_{N}$ on $\mathcal{A}\otimes\mathcal{M}_{N}(\mathbb{C})$ is positive for all $N\geq 1$, where $I_{N}$ denotes the identity map on the algebra $\mathcal{M}_{N}(\mathbb{C})$ of complex matrices of dimension $N$ .
\end{defin}

Though a direct application of this definition is always possible to check whether a given map is CP or not, the main tool in this sense is the famous Stinespring's theorem \cite{Sti55a}

\begin{theor}[Stinespring's Theorem]
For a map $\alpha:\mathcal{A}\to\mathcal{B}(\mathfrak{H})$ from a $C^{*}$ algebra to the space of bounded linear operators on the Hilbert space $\mathfrak{H}$, a sufficient and necessary condition for $\alpha$ to be CP is that there exists a representation $\pi:\mathcal{A}\to\mathfrak{K}$, $\mathfrak{K}$ a Hilbert space, such that
\begin{equation}
\alpha(X)=V\pi(X)V^{*}\quad\forall X\in\mathcal{A}\textrm{ and }V\in\mathcal{B}(\mathfrak{K})
\end{equation}
\end{theor}

This theorem is the base of the Kraus decomposition of a CP map \cite{Kra71a,Cho75a}, which is mostly used by physicists, and which we divide for future purposes into the cases of finite complex-matrices algebras and algebras of bounded linear operators on a Hilbert space.

\begin{theor}[Kraus decomposition]
%\begin{enumerate}
%\item[Finite algebras]
\emph{For finite algebras:} Let $\alpha:\mathcal{M}_{N}(\mathbb{C})\to\mathcal{M}_{N}(\mathbb{C})$ be a linear map. Then $\alpha$ is CP if and only if there exist $M_{1},\dots,M_{K}$ elements in $\mathcal{M}_{N}(\mathbb{C})$ such that
\begin{equation}
\alpha(X)=\sum_{p=1}^{K}M_{p}XM_{p}^{\dagger}\quad\forall X\in\mathcal{M}_{N}(\mathbb{C})
\end{equation}
%\item[Infinite algebras]

\emph{For infinite algebras:} Let $\alpha:\mathcal{B}(\mathfrak{H})\to\mathcal{B}(\mathfrak{H})$ be a linear map. Then $\alpha$ is CP if and only if there exists a sequence of operators $M_{1},M_{2},\dots,$, each one in $\mathcal{B}(\mathfrak{H})$ such that
\begin{equation}
\alpha(X)=\sum_{p=1}^{\infty}M_{p}XM_{p}^{\dagger}\quad\forall X\in\mathcal{B}(\mathfrak{H})
\end{equation}

The sum converges in the strong topology of $\mathcal{B}(\mathfrak{H})$.
%\end{enumerate}
\end{theor}

Note that the Kraus decomposition assures that a dynamical map is CP, but the problem then turns into finding the Kraus decomposition of an arbitrary linear map $\alpha$. This means that if one is not able to find one, we cannot conclude whether $\alpha$ is CP or not. Some results can be found in the literature \cite{BenFloRom02a} which reduces the complete positivity of a given map $\alpha$ to the positivity of the tensor map $\alpha\otimes\alpha$, which turns out to be also a complex question in practice. In the following sections, after providing a simplified proof of the Jamio\l kowski criterion for matrix algebras, once proven a map is CP, we give a procedure to build as many Kraus representations as desired.

\section{CP linear maps on $\mathcal{M}_{N}(\mathbb{C})$}
\label{Matrices}

We will focus on CP linear maps on complex matrices algebras, representing finite quantum systems. We make use of the following common notation. $\{E_{ij}\}_{i,j=1,\cdots,N}$ denotes the Weyl basis of $\mathcal{M}_{N}(\mathbb{C})$, i.e.\ those matrices with components $[E_{ij}]_{mn}=\delta_{im}\delta_{jn}$ or in physical notation $E_{ij}\equiv\ket{i}\bra{j}$. $\{e_{n}\}_{n=1,\cdots,N}$ will denote the basis on the complex vector space $\mathbb{C}^{N}$ and $(\cdot,\cdot)_{K}$  the standard complex scalar product which endows $\mathbb{C}^{K}$ with the well-known inner product space structure such that under these conventions we have $(e_{m},E_{ij}e_{n})_{N}=\delta_{mi}\delta_{jn}$ for all $i,j,m,n=1,\cdots,N$.\\
Generically we will denote by $\Lambda:\mathcal{M}_{N}(\mathbb{C})\to\mathcal{M}_{N}(\mathbb{C})$ a linear map from the algebra of complex matrices to itself. The set of such linear maps will be denoted as usual by $\mathcal{L}(\mathcal{M}_{N}(\mathbb{C}))$.

\begin{defin}
The map $\mathcal{J}_{e}:\mathcal{L}(\mathcal{M}(\mathbb{C^{N}}))\to\mathcal{M}_{N}(\mathbb{C})\otimes\mathcal{M}_{N}(\mathbb{C})$ is defined by

\begin{eqnarray}
\mathcal{J}_{e}&:\mathcal{L}(\mathcal{M}_{N}(\mathbb{C}))&\to\mathcal{M}_{N}(\mathbb{C})\otimes\mathcal{M}_{N}(\mathbb{C})\nonumber\\
&\Lambda&\to\mathcal{J}_{e}[\Lambda]=\sum_{i,j=1}^{N}\Lambda[E_{ij}]\otimes E_{ij}
\end{eqnarray}
This map is known as \emph{Jamio\l kowski isomorphism}.

\end{defin}
 
The subscript $e$ refers to the orthonormal basis in which the Weyl basis $\{E_{ij}\}$ is constructed. The main result is the proof of the criterion based upon this isomorphism, i.e.\ upon the so-called Jamio\l kowski criterion:

\begin{theor}[Jamio\l kowski criterion]
\label{TheMat}
Let $\Lambda:\mathcal{M}_{N}(\mathbb{C})\to\mathcal{M}_{N}(\mathbb{C})$ be a linear map. Then $\Lambda$ is CP if, and only if, $\mathcal{J}_{e}[\Lambda]\geq 0$.
\end{theor}

The proof we provide here rests upon the Kraus decomposition of a linear map.

%\begin{theor}
%the following conditions are satisfied

%\begin{enumerate}
%\item[(i)\label{HerCon}] Hermiticity:
%\begin{equation}
%\Lambda_{ijmn}=\Lambda_{jinm}^{*}
%\end{equation}
%\item[(ii)\label{PosDef}] PSD condition:\\
%The matrix with indices 
%\begin{equation}
%T_{[ij][mn]}\equiv \Lambda_{jnim}\quad i,j,m,n=1,\cdots,N
%\end{equation}

%must be positive semidefinite. 
%\end{enumerate}
%In particular $\Lambda_{nnmm}\geq 0$ for all $n,m=1,\cdots,N$ and if $\Lambda_{nnmm}=0$ for some $n,m\in\{1,\cdots,N\}$, then

%\begin{subequations}
%\begin{eqnarray}
%\Lambda_{njni}&=&0\ \forall i,j\in\{1,\cdots,N\}\\
%\Lambda_{imjm}&=&0\ \forall i,j\in\{1,\cdots,N\}
%\end{eqnarray}
%\end{subequations}
%\end{theor}

\begin{proof}

Firstly we will prove that if $\Lambda$ is CP, then $\mathcal{J}_{e}[\Lambda]\geq 0$. If $\Lambda$ is CP, then \cite{Cho75a} there exists $K$ elements $M_{k}\in\mathcal{M}_{N}(\mathbb{C})$ such that 

\begin{equation}
\Lambda[A]=\sum_{p=1}^{K}M_{p}AM_{p}^{\dagger}\quad \forall A\in\mathcal{M}_{N}(\mathbb{C})
\end{equation}

In particular, applying this Kraus decomposition to the Weyl matrices, one obtains

\begin{equation}\label{BasRel}
\sum_{p=1}^{K}M_{p}E_{ij}M_{p}^{\dagger}=\sum_{m,n=1}^{N}\Lambda_{ijmn}E_{mn}
\end{equation}

\noindent where $\Lambda_{ijmn}$ are the components of $\Lambda[E_{ij}]$ in the Weyl basis. Taking scalar products in \eqref{BasRel} one finds

\begin{subequations}

\begin{eqnarray}
\Lambda_{ijnm}&=&\sum_{p=1}^{K}(e_{n},M_{p}e_{i})_{N}(e_{j},M_{p}^{\dagger}e_{m})_{N}\nonumber\\
\label{ScaPro1}&=&\sum_{p=1}^{K}(e_{n},M_{p}e_{i})_{N}(e_{m},M_{p}e_{j})^{*}_{N}
\end{eqnarray}

\noindent where the Parseval relation has been used. Next, define $N^{2}$ $K$-dimensional complex vectors by $$f_{ij}\equiv\Big((e_{i},M_{1}e_{j})_{N},\cdots,(e_{i},M_{K}e_{j})_{N}\Big)\in\mathbb{C}^{K}$$

\noindent Relation \eqref{ScaPro1} can then be rewritten as

\begin{equation}\label{ScaPro2}
\Lambda_{ijnm}=(f_{mj},f_{ni})_{K}
\end{equation}

\noindent Since $(\cdot,\cdot)_{K}$ is the standard inner product in $\mathbb{C}^{K}$, from equation \eqref{ScaPro2} both Hermiticity and positive semidefiniteness of the matrix $\sum_{ij=1}^{N}\Lambda[E_{ij}]\otimes E_{ij}\in\mathcal{M}(\mathbb{C}^{N^{2}})$ follows from the Hermiticity and positive definiteness of the standard scalar product $(\cdot,\cdot)_{K}$. In other words, it follows from the positive semidefinite character of a matrix with structure $\left(\begin{smallmatrix}S&SC^{\dagger}\\
CS&CSC^{\dagger}
\end{smallmatrix}\right)$, where $S$ is a positive definite $N_{1}$-dimensional square matrix and $C$ is a $N_{1}\times(N^{2}-N_{1})$ rectangular matrix, with $N_{1}\equiv\textrm{rg}\left\{f_{ij}\right\}$, $i,j=1,\dots,N$. Hence (see appendix \ref{SemPos}) $\mathcal{J}_{e}[\Lambda]\geq 0$.\\

\noindent Conversely, if the linear map $\Lambda$ satisfies $\mathcal{J}_{e}[\Lambda]\geq 0$, i.e.\ the matrix $\sum_{ij=1}^{N}\Lambda[E_{ij}]\otimes E_{ij}$ is Hermitian and positive semidefiniteness, it adopts (see appendix \ref{SemPos}) the structure $\left(\begin{smallmatrix}S&SC^{\dagger}\\
CS&CSC^{\dagger}
\end{smallmatrix}\right)$ and then one can straightforwardly construct $N^{2}$ vectors $f_{ij}\in\mathbb{C}^{K}$, $i,j=1,\cdots,N$ such that relation $\eqref{ScaPro2}$ holds for all $i,j,m,n\in\{1,\cdots,N\}$.\\
Fix $K=N^{2}$. Locate those $m,n=1,\cdots,N$ such that $\Lambda_{nnmm}=0$, say $M<K$ of them (indices set $J$, for brevity). Then for these indices define \footnote{$[f_{ij}]$ denotes a vector whose components are the vectors $f_{ij}$ themselves.} $[f_{nm}]=C[f_{ij}]$ $i,j=1,\dots,N_{1}$. The problem is then reduced to find $K_{1}\equiv N^{2}-M$ vectors $\{f_{k}\}_{k=1,\cdots,N^{2}-M}$ such that the matrix of their scalar products is a given positive definite and Hermitian matrix. This can be accomplished using a similar method to the Gram-Schmidt orthonormalization process. An elementary proof is included in the form of a lemma in the appendix \ref{Lem} and thus is always possible. The reader should notice that since there exists an infinity of initial orthonormal bases (cf.\ appendix), one can also find an infinity of vectors satisfying the required conditions. This implies that the Kraus decomposition is not unique, as it is well-known.

\end{subequations}

\end{proof}

In this case of finite algebras, the use of matrices makes the application of the preceding theorem an extremely simple exercise. This is the content of the following theorem, which is but a reformulation of theorem \ref{TheMat}:

\begin{theor}\label{NecConMat}
Let $\Lambda:\mathcal{M}_{N}(\mathbb{C})\to\mathcal{M}_{N}(\mathbb{C})$ be a linear map. Then $\Lambda$ is CP if, and only if, the matrix $\Lambda_{W}$ defined below satisfies the following two properties:
\begin{enumerate}
\item[i.\label{HerMat}] $\Lambda_{W}=\Lambda_{W}^{\dagger}$.
\item[ii.\label{PosDefMat}] $\Lambda_{W}$ is positive semidefinite.
\end{enumerate}
In particular, if there is a zero element in the diagonal, then their corresponding row and column must be also zero. The matrix $\Lambda_{W}$ is defined by:

\[
\Lambda_{W}=\left(\begin{matrix}
\Lambda_{1111}&\dots&\Lambda_{111N}&\dots\dots&\Lambda_{N111}&\dots&\Lambda_{1N1N}\\
\vdots&\ddots &\vdots&\vdots&\vdots&\ddots&\vdots\\
\Lambda_{11N1}&\dots&\Lambda_{11NN}&\dots\dots&\Lambda_{1NN1}&\dots&\Lambda_{1NNN}\\
&\vdots&&\vdots&&\vdots&\\
\Lambda_{N111}&\dots&\Lambda_{N11N}&\dots\dots&\Lambda_{NN11}&\dots&\Lambda_{NN1N}\\
\vdots&\ddots &\vdots&\vdots&\vdots&\ddots&\vdots\\
\Lambda_{N1N1}&\dots&\Lambda_{N1NN}&\dots\dots&\Lambda_{NNN1}&\dots&\Lambda_{NNNN}
\end{matrix}\right)
\]
\end{theor}

\begin{proof}
It is the explicit expression of $\sum_{ij=1}^{N}\Lambda[E_{ij}]\otimes E_{ij}$ in the tensor product basis.\\%an elementary reorganization of the original $[\Lambda_{[ij][mn]}]^{[ij]=1,\cdots,N^{2}}_{[mn]=1,\cdots,N^{2}}$ matrix defining $\Lambda$ designed to check both properties at a glance.\\
However for completeness' sake we include the implication stated in theorem \ref{PosDefMat}. If a matrix is Hermitian and positive semidefinite, then there exists a diagonal matrix $D$ and a unitary matrix $P$ such that $M=P^{\dagger}DP$. The elements of $D$ are the eigenvalues of $M$ and the columns of $P$ the corresponding eigenvectors. Let $J$ denote the subset of indexes such that $d_{kk}=0$, $k\in J$. Let us also denote $\bar{J}=I-J$, where $I=\{1,\dots,N\}$, $N=\textrm{dim} M$. Then suppose $M_{jj}=0$, which implies

\begin{equation}
M_{jj}=\sum_{m\in I}|p_{mj}|^{2}d_{jj}=0\Rightarrow p_{mj}=0\quad \forall j\in\bar{J}
\end{equation}

We immediately conclude

\begin{subequations}
\begin{eqnarray}
M_{jk}&=&\sum_{m\in J}p_{mj}^{*}p_{mk}d_{mm}+\sum_{m\in\bar{J}}p_{mj}^{*}p_{mk}d_{mm}=0 \quad\forall k\in I\\
M_{kj}&=&\sum_{m\in J}p_{mk}^{*}p_{mj}d_{mm}+\sum_{m\in\bar{J}}p_{mk}^{*}p_{mj}d_{mm}=0 \quad\forall k\in I
\end{eqnarray}
\end{subequations}

Note that the converse is not always true.
\end{proof}

This proof, in conjuction with lemma \ref{Lem}, allows to extract the following corollaries:

\begin{corol}\label{CorKraMat}
Let $\Lambda$ be a CP map. Then $\mathcal{J}_{e}[\Lambda]=QQ^{\dagger}$ and the entries of the $k$th column of $Q^{*}$ are the components of the Kraus matrix $M_{k}$ in the Weyl basis $E_{ij}$. 
\end{corol}

\begin{corol}\label{AllSetKra}
Let $Q$ be as in corollary \ref{CorKraMat}. Then the entries of the $k$th column of $\tilde{Q}^{*}\equiv Q^{*}U$, where $U$ is a unitary $N^{2}$-dimensional matrix are the components of the Kraus matrix $\tilde{M}_{k}$ in the Weyl basis $E_{ij}$.
\end{corol}

These two corollaries can be summarized in words in a very simple way: any set of Kraus matrices of a CP map is given by the columns of a square root of $\mathcal{J}_{e}[\Lambda]$. Since there are infinite square roots, there will be infinite sets of Kraus representations, as already known. Finally, the number of Kraus matrices in such representations can also be readily obtained.

\begin{corol}
The number of matrices in a minimal Kraus representation of a CP linear map equals the number of positive eigenvalues of $\mathcal{J}_{e}[\Lambda]$.
\end{corol}

\begin{proof}
The number of matrices in a minimal Kraus representation equals the range of the set of Kraus matrices in any representation \cite{BucHor02a}. Since Kraus matrices are given by the columns of $Q^{*}$ and since eigenvectors are linearly independent, the number of non-null Kraus matrices will coincide with the number of eigenvector whose eigenvalue is non-null. 
\end{proof}

These tools will be illustrated with common well-known examples.

\section{Examples}
\label{Exa}
A very illustrative example of the immediate application of these results can be included by considering the transposition map $\Lambda[X]=X^{T}$. It is well-known that this is a non CP map, and to prove it one had traditionally to resort to consider maps $\Lambda\otimes I_{M}$ over the tensor product algebra $\mathcal{M}_{N}(\mathbb{C})\otimes\mathcal{M}_{M}(\mathbb{C})$ and study its positivity (cf.\ e.g.\ \cite{Arv69a}). With the preceding results, the task of deciding whether this is a CP map or not is elementary. Consider $\mathcal{M}_{2}(\mathbb{C})$ (nothing but computational complexity is gained in considering algebras of higher dimension). The matrix $\Lambda_{W}$ is easily obtained: 

\[
\Lambda_{W}=\left(
\begin{matrix}
1&0&0&0\\
0&0&1&0\\
0&1&0&0\\
0&0&0&1
\end{matrix}
\right)\]

It is clear that $\Lambda_{W}$ fails to satisfy the second condition in theorem \ref{NecConMat}, thus the transposition is not a CP map.\\% Note that this same proof is also valid for the map $\Lambda[X]=X^{\dagger}$.\\
Observe that once a linear map is proven to be CP, then we can also easily find one of its Kraus representations (indeed, we can find as many as we want). As an example of this technique, let us study in $\mathcal{M}_{2}(\mathbb{C})$ the linear map $\Lambda[X]=\frac{\lambda}{2}\textrm{tr}X I_{2}+\mu X$ with, in principle, $\lambda,\mu\in\mathbb{C}$. This map is especially relevant in quantum information theory \cite{ChuNie00a} and is known, under certain restrictions\footnote{Namely, the domain is not all $\mathcal{M}_{2}(\mathbb{C})$, but only the set of quantum states in which $X=X^{\dagger}$, $\textrm{tr}X=1$ and $X\geq 0$; also $\Lambda$ must be trace-preserving, i.e.\ $\lambda+\mu=1$ and $\lambda,\mu\in[0,1]$.}, as the depolarizing channel of one qubit. The matrix $\Lambda_{W}$ is 

\[
\Lambda_{W}=\left(
\begin{matrix}
\frac{\lambda}{2}+\mu&0&0&\mu\\
0&\frac{\lambda}{2}&0&0\\
0&0&\frac{\lambda}{2}&0\\
\mu&0&0&\frac{\lambda}{2}+\mu
\end{matrix}\right)\]

If $\Lambda_{W}$ is to be Hermitian, then both $\lambda$ and $\mu$ must be real numbers. If moreover it must also be positive semidefinite, then \footnote{The eigenvalues of $\Lambda_{W}$ are $\frac{\lambda}{2} \textrm{ (triple) and } \frac{\lambda}{2}+2\mu$.} $\lambda\geq 0$ and $\frac{\lambda}{2}+2\mu\geq 0$. The cases in which one of them is zero are trivial, so we shall focus on $\lambda,\mu\neq 0$. Since none of the eigenvalues is zero, we will need $4$ Kraus matrices. Following the preceding proof and the appendix, we can decompose $\Lambda_{W}=QQ^{\dagger}=PDP^{\dagger}$, with $P$ unitary and $D\geq 0$ diagonal, so that the columns of the matrix $Q^{*}\equiv P^{*}D^{1/2}$ contain the elements of each Kraus matrix:

\begin{eqnarray}
M_{1}&=&\sqrt{\frac{\lambda+4\mu}{2}}\mathbb{I}_{2}\\
M_{2}&=&\sqrt{\frac{\lambda}{4}}E_{12}\\
M_{3}&=&\sqrt{\frac{\lambda}{4}}E_{21}\\
M_{4}&=&\sqrt{\frac{\lambda}{4}}\sigma_{z}
\end{eqnarray}

\noindent where $\sigma_{z}$ is the $Z$ Pauli matrix. Notice that this is not the only set of Kraus matrices. We can make use of corollary \ref{AllSetKra} to find another set. In particular, choosing 

\begin{equation}
U=\left(\begin{matrix}
1&0&0&0\\
0&\frac{1}{\sqrt{2}}&-\frac{i}{\sqrt{2}}&0\\
0&\frac{1}{\sqrt{2}}&\frac{i}{\sqrt{2}}&0\\
0&0&0&1
\end{matrix}\right)
\end{equation}

\noindent we obtain another matrix $\tilde{Q}$, which produces the widely used set 

\begin{eqnarray}
\tilde{M}_{1}&=&\sqrt{\frac{\lambda+4\mu}{2}}\mathbb{I}_{2}\\
\tilde{M}_{2}&=&\sqrt{\frac{\lambda}{4}}\sigma_{x}\\
\tilde{M}_{3}&=&\sqrt{\frac{\lambda}{4}}\sigma_{y}\\
\tilde{M}_{4}&=&\sqrt{\frac{\lambda}{4}}\sigma_{z}
\end{eqnarray}

This technique shows the advantage of being possible to be systematically applied to any linear map in any dimension. Everything is reduced to a matter of computation.\\

\section{Conclusions}
\label{Conclu}

The main conclusion to be drawn is that to check the CP character of a linear map of matrix algebras, one only needs to check the Hermicitity and positive semidefiniteness of a properly constructed matrix. Besides, once complete positivity is assured, we also have a systematic method to find \emph{any} set of Kraus matrices of its Kraus representation.\\
The generalization to infinite algebras is currently under study. In this case, though the criterion is still valid, we must investigate the tractability of the method in these infinite cases.

\begin{acknowledgments}
We acknowledge financial support from the Spanish Ministry of Science and Technology through project FIS2004-01576. M.F. also acknowledges financial support from the University of Oviedo (ref.\ no.\ MB-04-514).
\end{acknowledgments}

\appendix

\section{}
\label{SemPos}

We will prove the following 

\begin{lemma}
A Hermitian $N^{2}$-dimensional matrix $M$ is positive semidefinite if, and only if, $M$ has the structure $\left(\begin{smallmatrix}S&SC^{\dagger}\\
CS&CSC^{\dagger}
\end{smallmatrix}\right)$, where $S$ is a Hermitian positive definite $N_{1}$-dimensional square matrix ($1\leq N_{1}\leq N^{2}$) and $C$ is a $N_{1}\times(N^{2}-N_{1})$ rectangular matrix.
\end{lemma}

\begin{proof}
$(\Rightarrow)$. Any Hermitian $N^{2}$-dimensional positive semidefinite matrix can be written as 

\begin{equation}
M=\left(\begin{smallmatrix}X&Y\\
Z&T
\end{smallmatrix}\right)\left(\begin{smallmatrix}D&0\\
0&0
\end{smallmatrix}\right)\left(\begin{smallmatrix}X&Y\\
Z&T
\end{smallmatrix}\right)^{\dagger}
\end{equation}

\noindent where $D$ is a positive definite $N_{1}$-dimensional matrix, $X$ and $T$ are unitary matrices, with dimensions $N_{1}$ and $N^{2}-N_{1}$, respectively, and $Y$ and $Z$ are null matrices. Then $M$ adopts the preceding structure after recognizing $S=XDX^{\dagger}$ and $C=ZX^{\dagger}(=0)$.

$(\Leftarrow)$.  Let us suppose that $M$ shows the preceding structure. We will prove that then it can be written as $M=Q^{\dagger}Q$, thus it is positive semidefinite. Since $S$ is positive definite, one can find an $N_{1}$-dimensional matrix $\xi$ such that $S=\xi^{\dagger}\xi$. A valid $Q$ can then be $\frac{1}{\sqrt{2}}\left(\begin{smallmatrix}\xi&\xi C^{\dagger}\\
V\xi&V\xi C^{\dagger}
\end{smallmatrix}\right)$, where $V$ is an arbitrary $[(N^{2}-N_{1})\times N_{1}]$-dimensional matrix such that $V^{\dagger}V=\mathbb{I}_{N_{1}}$. Note that this matrix, in the given conditions, always exists, since $N^{2}-N_{1}\geq N_{1}$ and then one can always find a matrix $V=\left(\begin{smallmatrix}\sqrt{1-\lambda}\mathbb{I}_{N_{1}}\\
b
\end{smallmatrix}\right)$ such that $b^{\dagger}b=\lambda\mathbb{I}_{N_{1}}$, where $b$ is an $(N^{2}-2N_{1})\times N_{1}$-dimensional matrix and $\lambda\in(0,1)$.  The Hermiticity is evident.
\end{proof}

\section{}
\label{Lem}
The problem is, given a positive definite Hermitian matrix $S\in\mathcal{M}_{N}(\mathbb{C})$, to find $N$ vectors $\{f_{j}\}_{j=1,\cdots,N}$ in $\mathbb{C}^{N}$ such that $(f_{i},f_{j})_{N}=S_{ij}$. This is always possible.

\begin{lemma}
Given a positive definite Hermitian matrix $[S_{ij}]$, $i,j=1,\cdots,N$, there exists a set of vectors $\{f_{k}\}_{k=1,\cdots,N}$ such that $(f_{i},f_{j})_{N}=S_{ij}$.
\end{lemma}

\begin{proof}
Consider an arbitrary orthonormal basis $\{b_{k}\}_{k=1,\cdots,N}$ (which can always be found with the orthonormalization Gram-Schmidt method). Since the matrix $S$ is positive definite and Hermitian, one can always find a unitary matrix $P$ and a diagonal positive matrix $D$ such $S=PDP^{\dagger}$. Define the matrix $Q\equiv PD^{1/2}$. Then one can elementary check that the set of vectors

\begin{equation}
f_{i}=\sum_{j=1}^{N}Q^{*}_{ij}b_{j}\quad i=1,\cdots,N
\end{equation}

\noindent satisfies the desired properties.
\end{proof}
%\bibliographystyle{apsrev}
%\bibliography{/home/david/Bibliography/Biblio}

\begin{thebibliography}{22}
\expandafter\ifx\csname natexlab\endcsname\relax\def\natexlab#1{#1}\fi
\expandafter\ifx\csname bibnamefont\endcsname\relax
  \def\bibnamefont#1{#1}\fi
\expandafter\ifx\csname bibfnamefont\endcsname\relax
  \def\bibfnamefont#1{#1}\fi
\expandafter\ifx\csname citenamefont\endcsname\relax
  \def\citenamefont#1{#1}\fi
\expandafter\ifx\csname url\endcsname\relax
  \def\url#1{\texttt{#1}}\fi
\expandafter\ifx\csname urlprefix\endcsname\relax\def\urlprefix{URL }\fi
\providecommand{\bibinfo}[2]{#2}
\providecommand{\eprint}[2][]{\url{#2}}

\bibitem[{\citenamefont{Stinespring}(1955)}]{Sti55a}
\bibinfo{author}{\bibfnamefont{W.}~\bibnamefont{Stinespring}},
  \bibinfo{journal}{Proc. Amer. Math. Soc.} \textbf{\bibinfo{volume}{6}},
  \bibinfo{pages}{211} (\bibinfo{year}{1955}).

\bibitem[{\citenamefont{rmer}(1963)}]{Sto63a}
\bibinfo{author}{\bibfnamefont{E.} \bibnamefont{St\o rmer}},
  \bibinfo{journal}{Acta Math.} \textbf{\bibinfo{volume}{110}},
  \bibinfo{pages}{223} (\bibinfo{year}{1963}).

\bibitem[{\citenamefont{Arverson}(1969)}]{Arv69a}
\bibinfo{author}{\bibfnamefont{W.}~\bibnamefont{Arverson}},
  \bibinfo{journal}{Acta Math.} \textbf{\bibinfo{volume}{123}},
  \bibinfo{pages}{141} (\bibinfo{year}{1969}).

\bibitem[{\citenamefont{Choi}(1982)}]{Cho82a}
\bibinfo{author}{\bibfnamefont{M.-D.} \bibnamefont{Choi}},
  \bibinfo{journal}{Proc. Symp. Pure Math.} \textbf{\bibinfo{volume}{38}},
  \bibinfo{pages}{583} (\bibinfo{year}{1982}).

\bibitem[{\citenamefont{Peres}(1996)}]{Per96a}
\bibinfo{author}{\bibfnamefont{A.}~\bibnamefont{Peres}},
  \bibinfo{journal}{Phys. Rev. Lett.} \textbf{\bibinfo{volume}{77}},
  \bibinfo{pages}{1413} (\bibinfo{year}{1996}).

\bibitem[{\citenamefont{Horodecki et~al.}(1996)\citenamefont{Horodecki,
  Horodecki, and Horodecki}}]{HorHorHor96a}
\bibinfo{author}{\bibfnamefont{M.}~\bibnamefont{Horodecki}},
  \bibinfo{author}{\bibfnamefont{P.}~\bibnamefont{Horodecki}},
  \bibnamefont{and}
  \bibinfo{author}{\bibfnamefont{R.}~\bibnamefont{Horodecki}},
  \bibinfo{journal}{Phys. Lett. A} \textbf{\bibinfo{volume}{223}},
  \bibinfo{pages}{8} (\bibinfo{year}{1996}).

\bibitem[{\citenamefont{Lewenstein et~al.}(2000)\citenamefont{Lewenstein,
  Bruss, Cirac, Kraus, Samsonowicz, Sanpera, and
  Tarrach}}]{LewBruCirKraKusSamSanTar00a}
\bibinfo{author}{\bibfnamefont{M.}~\bibnamefont{Lewenstein}},
  \bibinfo{author}{\bibfnamefont{D.}~\bibnamefont{Bruss}},
  \bibinfo{author}{\bibfnamefont{J.}~\bibnamefont{Cirac}},
  \bibinfo{author}{\bibfnamefont{B.}~\bibnamefont{Kraus}},
  \bibinfo{author}{\bibfnamefont{J.}~\bibnamefont{Samsonowicz}},
  \bibinfo{author}{\bibfnamefont{A.}~\bibnamefont{Sanpera}}, \bibnamefont{and}
  \bibinfo{author}{\bibfnamefont{R.}~\bibnamefont{Tarrach}},
  \bibinfo{journal}{quant-ph/0006064}  (\bibinfo{year}{2000}).

\bibitem[{\citenamefont{Bruss et~al.}(2001)\citenamefont{Bruss, Cirac,
  Horodecki, Hulpke, Kraus, Lewenstein, and
  Sanpera}}]{BruCirHorHulKraLewSan01a}
\bibinfo{author}{\bibfnamefont{D.}~\bibnamefont{Bruss}},
  \bibinfo{author}{\bibfnamefont{J.}~\bibnamefont{Cirac}},
  \bibinfo{author}{\bibfnamefont{P.}~\bibnamefont{Horodecki}},
  \bibinfo{author}{\bibfnamefont{F.}~\bibnamefont{Hulpke}},
  \bibinfo{author}{\bibfnamefont{B.}~\bibnamefont{Kraus}},
  \bibinfo{author}{\bibfnamefont{M.}~\bibnamefont{Lewenstein}},
  \bibnamefont{and} \bibinfo{author}{\bibfnamefont{A.}~\bibnamefont{Sanpera}},
  \bibinfo{journal}{quant-ph/0110081}  (\bibinfo{year}{2001}).

\bibitem[{\citenamefont{Davies}(1976)}]{Dav76a}
\bibinfo{author}{\bibfnamefont{E.}~\bibnamefont{Davies}},
  \emph{\bibinfo{title}{{Q}uantum {T}heory of {O}pen {S}ystems}}
  (\bibinfo{publisher}{Academic Press}, \bibinfo{address}{London},
  \bibinfo{year}{1976}).

\bibitem[{\citenamefont{Lindblad}(1976)}]{Lin76a}
\bibinfo{author}{\bibfnamefont{G.}~\bibnamefont{Lindblad}},
  \bibinfo{journal}{Commun. Math. Phys.} \textbf{\bibinfo{volume}{48}},
  \bibinfo{pages}{119} (\bibinfo{year}{1976}).

\bibitem[{\citenamefont{Gorini et~al.}(1976)\citenamefont{Gorini, Kossakowski,
  and Sudarshan}}]{GorKosSud76a}
\bibinfo{author}{\bibfnamefont{V.}~\bibnamefont{Gorini}},
  \bibinfo{author}{\bibfnamefont{A.}~\bibnamefont{Kossakowski}},
  \bibnamefont{and}
  \bibinfo{author}{\bibfnamefont{E.}~\bibnamefont{Sudarshan}},
  \bibinfo{journal}{J. Math. Phys.} \textbf{\bibinfo{volume}{17}},
  \bibinfo{pages}{821} (\bibinfo{year}{1976}).

\bibitem[{\citenamefont{Alicki and Lendi}(1987)}]{AliLen87a}
\bibinfo{author}{\bibfnamefont{R.}~\bibnamefont{Alicki}} \bibnamefont{and}
  \bibinfo{author}{\bibfnamefont{K.}~\bibnamefont{Lendi}},
  \emph{\bibinfo{title}{{Q}uantum {D}ynamical {S}emigroups and
  {A}pplications}}, Lecture Notes in Physics 286
  (\bibinfo{publisher}{Springer-Verlag}, \bibinfo{address}{Berlin},
  \bibinfo{year}{1987}).

\bibitem[{\citenamefont{Breuer and Petruccione}(2002)}]{BrePet02a}
\bibinfo{author}{\bibfnamefont{H.}~\bibnamefont{Breuer}} \bibnamefont{and}
  \bibinfo{author}{\bibfnamefont{F.}~\bibnamefont{Petruccione}},
  \emph{\bibinfo{title}{{T}he {T}heory of {O}pen {Q}uantum {S}ystems}}
  (\bibinfo{publisher}{Oxford University Press}, \bibinfo{address}{Oxford},
  \bibinfo{year}{2002}).

\bibitem[{\citenamefont{Buchleitner and Hornberger}(2002)}]{BucHor02a}
\bibinfo{editor}{\bibfnamefont{A.}~\bibnamefont{Buchleitner}} \bibnamefont{and}
  \bibinfo{editor}{\bibfnamefont{K.}~\bibnamefont{Hornberger}}, eds.,
  \emph{\bibinfo{title}{{C}oherent {E}volution in {N}oisy {E}nvironments}},
  Lecture Notes in Physics 611 (\bibinfo{publisher}{Springer},
  \bibinfo{address}{Berlin}, \bibinfo{year}{2002}).

\bibitem[{\citenamefont{Kraus}(1971)}]{Kra71a}
\bibinfo{author}{\bibfnamefont{K.}~\bibnamefont{Kraus}}, \bibinfo{journal}{Ann.
  Phys.} \textbf{\bibinfo{volume}{64}}, \bibinfo{pages}{311}
  (\bibinfo{year}{1971}).

\bibitem[{\citenamefont{Kraus}(1983)}]{Kra83a}
\bibinfo{author}{\bibfnamefont{K.}~\bibnamefont{Kraus}},
  \emph{\bibinfo{title}{{S}tates, {E}ffects and {O}perations: {F}undamental
  {N}otions of {Q}uantum {T}heory}} (\bibinfo{publisher}{Springer},
  \bibinfo{address}{Berlin}, \bibinfo{year}{1983}).

\bibitem[{\citenamefont{de~Pillis}(1967)}]{Pil67a}
\bibinfo{author}{\bibfnamefont{J.}~\bibnamefont{de~Pillis}},
  \bibinfo{journal}{Pac. J. Math.} \textbf{\bibinfo{volume}{23}},
  \bibinfo{pages}{129} (\bibinfo{year}{1967}).

\bibitem[{\citenamefont{Jamio\l kowski}(1972)}]{Jam72a}
\bibinfo{author}{\bibfnamefont{A.}~\bibnamefont{Jamio\l kowski}},
  \bibinfo{journal}{Rep. Math. Phys.} \textbf{\bibinfo{volume}{3}},
  \bibinfo{pages}{275} (\bibinfo{year}{1972}).

\bibitem[{\citenamefont{Choi}(1975)}]{Cho75a}
\bibinfo{author}{\bibfnamefont{M.-D.} \bibnamefont{Choi}},
  \bibinfo{journal}{Lin. Alg. and Appl.} \textbf{\bibinfo{volume}{10}},
  \bibinfo{pages}{285} (\bibinfo{year}{1975}).

\bibitem[{\citenamefont{Choi}(1972)}]{Cho72a}
\bibinfo{author}{\bibfnamefont{M.-D.} \bibnamefont{Choi}},
  \bibinfo{journal}{Can. J. Math.} \textbf{\bibinfo{volume}{24}},
  \bibinfo{pages}{520} (\bibinfo{year}{1972}).

\bibitem[{\citenamefont{Benatti et~al.}(2002)\citenamefont{Benatti, Floreanini,
  and Romano}}]{BenFloRom02a}
\bibinfo{author}{\bibfnamefont{F.}~\bibnamefont{Benatti}},
  \bibinfo{author}{\bibfnamefont{R.}~\bibnamefont{Floreanini}},
  \bibnamefont{and} \bibinfo{author}{\bibfnamefont{R.}~\bibnamefont{Romano}},
  \bibinfo{journal}{J. Phys. A} \textbf{\bibinfo{volume}{35}},
  \bibinfo{pages}{L551} (\bibinfo{year}{2002}).

\bibitem[{\citenamefont{Nielsen and Chuang}(2000)}]{ChuNie00a}
\bibinfo{author}{\bibfnamefont{M.A.}~\bibnamefont{Nielsen}} \bibnamefont{and}
  \bibinfo{author}{\bibfnamefont{I.L.}~\bibnamefont{Chuang}},
  \emph{\bibinfo{title}{{Q}uantum {C}omputation and {Q}uantum {I}nformation}}
  (\bibinfo{publisher}{Cambridge University Press},
  \bibinfo{address}{Cambridge}, \bibinfo{year}{2000}).

\end{thebibliography}

\end{document}